\documentclass[5p,number]{elsarticle}
\usepackage{amssymb,graphicx,hyperref}

\newcounter{bla}

\journal{Computer Physics Communications}

\begin{document}

\begin{frontmatter}
\title{Package \texttt{feyn.gle} for drawing Feynman diagrams, version 1.2}
\author{Andrey Grozin}
\ead{A.G.Grozin@inp.nsk.su}
\affiliation{organization={Budker Institute of Nuclear Physics},
addressline={Lavrentiev st. 10},
postcode={630090},
city={Novosibirsk},
country={Russia}}
\affiliation{organization={Bogoliubov Laboratory of Theoretical Physics,
Joint Institute for Nuclear Research},
postcode={141980},
city={Dubna},
country={Russia}}
\begin{abstract}
As compared to the previous version, there is one backward-incompatible change
and many new kinds of lines and other objects.
\end{abstract}
\end{frontmatter}

\textbf{Program summary}
\begin{small}
\noindent
\emph{Program Title:} feyn.gle\\
\emph{CPC Library link to program files:} (to be added by Technical Editor)\\
\emph{Developer's repository link:}\\
\url{https://github.com/andreygrozin/feyn.gle}\\
\emph{Licensing provisions:} BSD 3-clause\\
\emph{Programming language:} GLE~[1]\\
\emph{Journal reference of previous version:} A.~Grozin, Drawing Feynman diagrams with GLE,
Comput.\ Phys.\ Comm.\ 283 (2023) 108590\\
\emph{Does the new version supersede the previous version?:} Yes\\
\emph{Summary of revisions:} 1 incompatible change, many new kinds of lines and other objects\\

\end{small}

\section{Introduction}
\label{S:Intro}

\begin{sloppypar}
This paper describes a new version of the package \verb|feyn.gle|~\cite{Grozin:2022fde}
written in the language \verb|GLE|.
\end{sloppypar}

Programs for drawing Feynman diagrams can be subdivided into script-based ones and GUI-based ones.
In the case of a script-based program, a user has to write a script in some language to draw a diagram.
This approach has some advantages:
it is easy to copy a script from some earlier work and re-use it for a new work, maybe, with some adjustments;
it is easy to have a set of diagrams in exactly the same style:
same dimensions, identical subdiagrams, etc.;
script texts can be automatically generated by some programs.

\begin{sloppypar}
There are several programs where diagram scripts are written directly inside a \LaTeX{} file.
One of them is \verb|axodraw|~\cite{Vermaseren:1994je,Collins:2016aya}
(\url{https://ctan.org/pkg/axodraw}, \url{https://ctan.org/pkg/axodraw2}).
The first version requires the user to use plain \verb|latex| and go via \verb|dvi| and \verb|PostScript|;
the second version allows one to use \verb|pdflatex| or \verb|lualatex|,
but it is necessary to use also a separate program \verb|axohelp|.
\end{sloppypar}

\begin{sloppypar}
The package \verb|feynMF|~\cite{Ohl:1995kr} (\url{https://ctan.org/pkg/feynmf}) uses \verb|METAFONT| to generate bitmaps,
its variant \verb|feynMP| uses \verb|METAPOST| to generate \verb|PostScript|.
The package \verb|TikZ-Feynman|~\cite{Ellis:2016jkw} (\url{https://ctan.org/pkg/tikz-feynman})
uses \verb|PGF/TikZ| and can be used with both \verb|latex| and \verb|pdflatex|.
These packages can calculate vertex coordinates automatically.
However, tuning this automatic process to produce diagrams which look exactly as the author wants
can be a tedious and time-consuming process.
\end{sloppypar}

In the case of \verb|feyn.gle|, diagram scripts are written not in \verb|.tex| files
but in separate \verb|.gle| files.
This approach has both advantages and disadvantages.
On the one hand, there are more files, which have to be processed in some way;
it is also necessary to know basics of the \verb|GLE| language
(but it is very simple, similar to old versions of \verb|basic|).
On the other hand, coordinates of vertices can be given by not-quite-trivial formulas
from analytical geometry, something like $\sqrt{x^2+y^2}\sin(\alpha)$;
such formulas can be trivially written in \verb|GLE|,
but writing them in a \LaTeX{} source is troublesome.
It is trivial to produce many graphical formats from a \verb|GLE| script,
both vector (\verb|eps|, \verb|ps|, \verb|pdf|, \verb|svg|) and bitmap (\verb|png|, \verb|jpg|).
The author can, from a single \verb|GLE| script, produce a \verb|pdf| (or \verb|eps|) for including in a \LaTeX{} article,
an \verb|svg| for adding it to a web page,
a \verb|jpg| for including in a \verb|PowerPoint| (or similar) presentation, etc.

\begin{sloppypar}
Let's also briefly mention some GUI-based programs.
\verb|JaxoDraw|~\cite{Binosi:2003yf,Binosi:2008ig} (\url{https://jaxodraw.sourceforge.io/})
is a GUI \verb|Java| program.
A user creates a diagram in this GUI (this process is easy),
then the program can export a \LaTeX{} source using \verb|axodraw|
(it can also export several graphics formats directly).
Another GUI \verb|Java| program is
\verb|FeynGame|~\cite{Harlander:2020cyh,Harlander:2024qbn,Bundgen:2025utt}
(\url{https://web.physik.rwth-aachen.de/user/harlander/software/feyngame/}).
It knows what propagators and vertices are present in specific quantum field theory models.
\end{sloppypar}

\section{Non-planar diagrams}
\label{S:NonPlan}

Some authors prefer to mark all vertices in the diagram (using \verb|@Vert|);
then intersections of lines without it are \emph{not} vertices.
Other authors prefer to have small gaps in lower lines
so that upper lines go through these gaps.
In the previous versions this was done by the subroutine \verb|Shadow|.
It produced unnecessary complicated structure of \verb|pdf| files,
and did not work correctly if the current line had some parameters
that are automatically reset to defaults after the call.
In the new version there is a single backward-incompatible change:
\verb|Shadow| is now absent.
Instead, one can set the variable \verb|Gap = |$f$ ($f>0$)
before drawing a line of any kind.
Then, before actual drawing, at the fraction $f$ of the line length
a small area of the length $2*\mathtt{GapL}$ and the width $2*\mathtt{GapA}$
will be filled with the color \verb|GapC$| (by default white).
After drawing the line, \verb|Gap| is set to its default value $0$.
For example,
\small
\begin{verbatim}
x = 1.5
y = 0.75*x
a = 0.5
r = (x^2+y^2)/(2*y)
x1 = 0.6
y1 = sqrt(r^2-x1^2)-r+y
amove -x-a 0
@Fermion -x 0
@Fermion2 0 y x 0
@Fermion2 0 -y -x 0
amove x 0
@Fermion x+a 0
amove x1 -y1
@Fermion -x1 y1
amove -x1 -y1
Gap = 0.5
@Fermion x1 y1
\end{verbatim}
\normalsize
\includegraphics{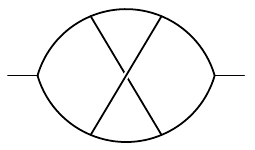}

The parameters \verb|GapL|, \verb|GapA|, \verb|GapC$| can be changed:
\small
\begin{verbatim}
amove -0.5*x -y
@Fermion -0.5*x y
amove 0 -y
@Fermion 0 y
amove 0.5*x -y
@Fermion 0.5*x y
amove -x 0
GapC$ = "gray10"
Gap = 0.25
@Feyn x 0
amove -x 0
GapL = 2*DefGapL
Gap = 0.5
@Feyn x 0
GapL = DefGapL
GapA = 2*DefGapA
Gap = 0.75
@Fermion x 0
GapA = DefGapA
\end{verbatim}
\normalsize
\includegraphics{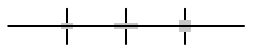}\\
We remind the reader that \verb|@Feyn x y| draws an invisible line
from the current point to $(x,y)$;
this trick can be used to make gaps at intersections with several lower lines.
Alternatively, one can set a sufficiently large \verb|GapL|:
\small
\begin{verbatim}
amove -0.5*x 0
@Fermion -0.5*x y+a
amove 0 0
@Fermion 0 y+a
amove 0.5*x 0
@Fermion 0.5*x y+a
amove -x 0
GapC$ = "gray10"
GapL = 1
Gap = 0.5
@Fermion2 0 y x 0
\end{verbatim}
\normalsize
\includegraphics{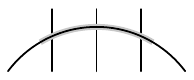}

Of course, all lines can be of any kinds;
if some lower lines (in which we cut gaps) are wide (e.g., photon or gluon),
one may want to increase \verb|GapA|.

One more minor backward-incompatible change:
the parameter \verb|ArrowD| has been removed (it was never used in practice).

\section{New kinds of lines}
\label{S:Lines}

Dotted lines existed in previous versions;
in this new version they have 2 new control parameters.
The gap between the beginning of a line and the first dot
is \verb|DotsB| times the normal distance between dots;
similarly, \verb|DotsE| controls the gap at the end of the line:
\small
\begin{verbatim}
amove 0 -0.5*y
@Fermion 0 3.5*y
amove x -0.5*y
@Fermion x 3.5*y
amove 0 0
@Dots x 0
amove 0 y
DotsB = 0
DotsE = 0
@Dots x y
amove 0 2*y
DotsE = 1
@Dots x 2*y
amove 0 3*y
DotsB = 0.5
DotsE = 0.5
@Dots x 3*y
\end{verbatim}
\normalsize
\includegraphics{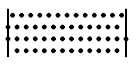}\\
By default, both parameters are 1,
thus reproducing the behavior of previous versions.

\subsection{Double lines}
\label{S:Double}

In addition to normal double lines,
there are double variants of dashed, dotted, zigzag, photon lines.
The distance between the lines is $2*\mathtt{DoubleA}$.
Double dashed lines support all control parameters of dashed lines
(\verb|DashL|, \verb|DashF|, \verb|DashN|, \verb|DashB|, \verb|DashE|):
\small
\begin{verbatim}
amove 0 -0.5*y
@Fermion 0 4.5*y
amove x -0.5*y
@Fermion x 4.5*y
amove 0 0
@DDash x 0
amove 0 y
DashB = -1
DashE = -1
@DDash x y
amove 0 2*y
DashB = 1
@DDash x 2*y
amove 0 3*y
DashB = 0.5
DashE = 0.5
@DDash x 3*y
amove 0 4*y
DashB = -0.5
DashE = -0.5
@DDash x 4*y
\end{verbatim}
\normalsize
\includegraphics{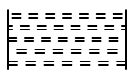}\\
At the same time they support the control parameters
\verb|DoubleB|, \verb|DoubleE|, \verb|DoubleA| of \verb|Double| lines
(statements similar to the previous example are omitted for brevity):
\small
\begin{verbatim}
amove 0 0
@DDash x 0
amove 0 y
DoubleB = 1
DoubleE = 1
@DDash x y
amove 0 2*y
DoubleB = -1
DoubleE = -1
@DDash x 2*y
amove 0 3*y
DoubleB = 1
DoubleE = -1
@DDash x 3*y
amove 0 4*y
DoubleB = -1
DoubleE = 1
@DDash x 4*y
\end{verbatim}
\normalsize
\includegraphics{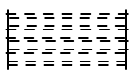}

Double dotted lines support all control parameters of dotted lines
(\verb|DotR|, \verb|DotsL|, \verb|FillC$|, \verb|DotsB|, \verb|DotsE|)
and parameters of double lines (\verb|DoubleA|, \verb|DoubleB|, \verb|DoubleE|).
In order to avoid overlapping of dots one may want to increase \verb|DoubleA|
or to decrease \verb|DotR| (or both).
\small
\begin{verbatim}
DotR = 0.75*DefDotR
DoubleA = 1.25*DefDoubleA
amove 0 0
@DDots x 0
amove 0 y
DotsB = 0
DotsE = 0
@DDots x y
amove 0 2*y
DotsE = 1
@DDots x 2*y
amove 0 3*y
DotsB = 0.5
DotsE = 0.5
@DDots x 3*y
\end{verbatim}
\normalsize
\includegraphics{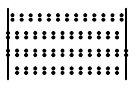}
\small
\begin{verbatim}
DotsB = 0
DotsE = 0
amove 0 0
@DDots x 0
amove 0 y
DoubleB = 1
DoubleE = 1
@DDots x y
amove 0 2*y
DoubleB = -1
DoubleE = -1
@DDots x 2*y
amove 0 3*y
DoubleB = 1
DoubleE = -1
@DDots x 3*y
amove 0 4*y
DoubleB = -1
DoubleE = 1
@DDots x 4*y
\end{verbatim}
\normalsize
\includegraphics{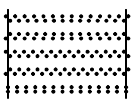}

Double zigzag lines support all control parameters of zigzag lines
(\verb|PhotonL|, \verb|PhotonA|, \verb|PhotonN|) and \verb|DoubleA|,
but \emph{not} \verb|DoubleB|, \verb|DoubleE|
(this would produce weird results).
\small
\begin{verbatim}
amove 0 0
@DZigzag x 0
amove 0 y
PhotonA = -DefPhotonA
@DZigzag x y
PhotonA = DefPhotonA
amove 0 2*y
PhotonN = -1
@DZigzag x 2*y
\end{verbatim}
\normalsize
\includegraphics{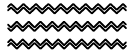}\\
Double photon lines are similar.
\small
\begin{verbatim}
amove 0 0
@DPhoton x 0
amove 0 y
PhotonA = -DefPhotonA
@DPhoton x y
PhotonA = DefPhotonA
amove 0 2*y
PhotonN = -1
@DPhoton x 2*y
\end{verbatim}
\normalsize
\includegraphics{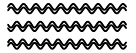}

\subsection{More new features}
\label{S:More}

\begin{sloppypar}
Combinations of zigzag, photon, gluon lines with fermion ones
are sometimes used for superpartners --- wino, zino, photino, gluino.
\end{sloppypar}
\small
\begin{verbatim}
amove 0 0
@FZigzag x 0
amove 0 y
@FPhoton x y
amove 0 2*y
@FGluon x 2*y
\end{verbatim}
\normalsize
\includegraphics{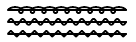}\\
\verb|FZigzag|, \verb|FPhoton| support all control parameters of photon lines;
\verb|FGluon| --- those of gluon lines.

In addition to ordinary vertices --- filled circles of radius \verb|DotR|,
the new version supports vertices with the shape of regular polygons.
Their size, i.e.\ the distance from the center to a polygon vertex, is given by \verb|VertA|.
If you don't tune such an $n$-gon, one of its vertices looks upward.
If you set \verb|VertR|${}=\xi$, it will be rotated counterclockwise
by the angle $\pi\xi/n$
(i.e., $\xi=2$ is an element of the symmetry group and changes nothing;
$\xi=1$ produces a polygon with a side looking upward).
\small
\begin{verbatim}
amove 0 0
@Fermion 2*x 0
amove x 0
@PVert 3
@Photon x x
amove 2*x+a 0
@Fermion 4*x+a 0
amove 3*x+a 0
VertR = 1
@PVert 3
@Photon 2*x+a x
amove 3*x+a 0
PhotonA = -PhotonA
@Photon 4*x+a x
PhotonA = -PhotonA
amove 0 x+a
@Fermion 2*x x+a
amove x x+a
@PVert 4
@Photon x 2*x+a
amove 2*x+a x+a
@Fermion 4*x+a x+a
amove 3*x+a x+a
VertR = 1
@PVert 4
@Photon 2*x+a 2*x+a
amove 3*x+a x+a
PhotonA = -PhotonA
@Photon 4*x+a 2*x+a
PhotonA = -PhotonA
\end{verbatim}
\normalsize
\includegraphics{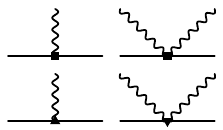}\\
\verb|GLE| markers can also be used as vertices.

\verb|LDot f| puts a dot at the fraction $f$ of the length of a previous line.
It is useful, e..g., for drawing master integrals with some dots on lines.
\small
\begin{verbatim}
amove -x 0
@Fermion2 0 y x 0
@Ldot 1/3
@Ldot 2/3
\end{verbatim}
\normalsize
\includegraphics{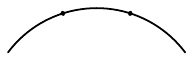}

\section{Pattern lines}
\label{S:Pattern}

This version introduces a new kind of lines which generalizes dashed and dotted lines.
They are built as a sequence of patterns, each of them has the length \verb|PatternL|
and contains some dashes and/or dots.

Unfortunately, there are no arrays in the language \verb|GLE|,
and we have to emulate them by sets of plain variables.
To set up dashes in the pattern a user defines an ``array''\\[2mm]
\verb|DashPattern = |$n$\\
\verb|DashPattern1 = |$\xi_1$\\
\verb|DashPattern2 = |$\xi_2$\\
\ldots\\
\verb|DashPattern|$n$\verb| = |$\xi_n$\\[2mm]
Important!
$n$ must be a positive integer;
all elements between \verb|DashPattern1| and \verb|DashPattern|$n$ must be set, without omissions.
Otherwise runtime errors will happen,
the program does not check that these rules are hold.

In the case of a \verb|DashPattern|, the values $\xi_i$ must satisfy
\[
0 \le \xi_1 < \xi_2 < \cdots < \xi_n < 1\,.
\]
They are fractions of the period (the length of a single pattern).
The program produces a dash from 0 to $\xi_1$,
then a gap to $\xi_2$,
then a dash to $\xi_3$,
and so on.
The parameter \verb|PatternB = |$\xi$ ($0\le\xi<1$) says that the beginning of the line
is at the fraction $\xi$ of the pattern;
\verb|PatternE| controls the end of the line in the same way.
The parameter \verb|PatternN| changes the number of periods (patterns) on a line.
\small
\begin{verbatim}
DashPattern = 3
DashPattern1 = 0.4
DashPattern2 = 0.6
DashPattern3 = 0.8
amove 0 0
@Pattern x 0
amove 0 y
PatternE = 0.4
@Pattern x y
amove 0 2*y
PatternB = 0.6
PatternE = 0.8
@Pattern x 2*y
\end{verbatim}
\normalsize
\includegraphics{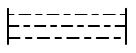}

Don't forget that \emph{all} kinds of lines have variants
which produce straight lines, arcs, or circles.
For example,
\small
\begin{verbatim}
amove -x 0
PatternE = 0.4
@Pattern2 0 y x 0
\end{verbatim}
\normalsize
\includegraphics{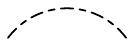}
\small
\begin{verbatim}
amove -x 0
@PatternC 0 0
\end{verbatim}
\normalsize
\includegraphics{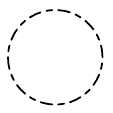}

\begin{sloppypar}
\verb|DPattern| produces a double pattern line.
It supports all control parameters of \verb|Pattern|
and also \verb|DoubleA|, \verb|DoubleB|, \verb|DoubleE|.
\end{sloppypar}
\small
\begin{verbatim}
PatternE = 0.4
amove 0 0
@DPattern x 0
amove 0 y
DoubleB = 1
DoubleE = 1
@DPattern x y
amove 0 2*y
DoubleB = 1
DoubleE = -1
@DPattern x 2*y
\end{verbatim}
\normalsize
\includegraphics{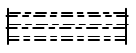}

If \verb|DashPattern| is 0, do dashes are drawn.
A user can define a dots pattern --- an ``array'' with positions of dots.
\small
\begin{verbatim}
DotsPattern = 2
DotsPattern1 = 0
DotsPattern2 = 1/3
amove 0 0
@Pattern x 0
amove 0 y
PatternE = 1/3
@Pattern x y
amove 0 2*y
PatternB = 1/3
PatternE = 0
@Pattern x 2*y
\end{verbatim}
\normalsize
\includegraphics{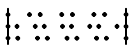}\\
And what is this rightmost dot on the lowest line?
Logically, there is an infinite line with an infinite sequence of dots patterns;
we see its part through a finite window.
This dot is the dot at $\xi=0$ (the first element of our ``array'')
of the \emph{next} period after the end of our line.
\verb|Pattern| does not consider this next period (it is outside our finite line)
except when it contains a dot at $\xi=0$ and \verb|PatternE|${}=0$.
Such situation cannot happen at the beginning of the line:
the \emph{previous} period (pattern) cannot have a dot at $\xi=1$
(all $\xi_i$ must be $<1$).
If \verb|DotsPatternE|${}\le0$ then a dot,
if logically present exactly at the end of the line, is not drawn.
Similarly, \verb|DotsPatternB| controls a (possible) dot exactly at the beginning of the line.
By default, both of these parameters are 1, and such dots are drawn.
By switching one of them (or both) to 0
we remove all dots in the above example which are on one (or both) of vertical lines.

\begin{sloppypar}
We can have both ``arrays'' \verb|DashPattern| and \verb|DotsPattern| together,
and produce various dashed-dotted lines.
\end{sloppypar}
\small
\begin{verbatim}
DashPattern = 1
DashPattern1 = 0.5
DotsPattern = 1
DotsPattern1 = 0.75
amove 0 -0.5*y
@Fermion 0 2.5*y
amove x -0.5*y
@Fermion x 2.5*y
amove 0 0
@Pattern x 0
amove 0 y
PatternE = 0.5
@Pattern x y
amove 0 2*y
PatternB = 0.75
PatternE = 0.75
@Pattern x 2*y
\end{verbatim}
\normalsize
\includegraphics{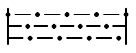}\\
Of course, double pattern lines \verb|DPattern| can use any combinations
of a dash pattern and a dot one;
it always support \verb|DoubleA|, \verb|DoubleB|, \verb|DoubleE|.

\section{Conclusion}
\label{S:Conc}

Let's summarize the differences between the version 1.1.3 described in~\cite{Grozin:2022fde}
and the new version 1.2.0 described here.

Backward-incompatible changes:
\begin{itemize}
\item The subroutines \verb|Shadow|, \verb|Shadow2|,
as well as their control parameters \verb|ShadowL|, \verb|ShadowA|, \verb|ShadowC$|,
have been removed.
Instead, use the control parameter \verb|Gap| with any kind of line,
and \verb|GapL|, \verb|GapA|, \verb|GapC$| for fine-tuning.
\item The control parameter \verb|ArrowD| has been removed
(it was useless from the very beginning).
\end{itemize}

New kinds of lines:
\begin{itemize}
\item Double dashed lines \verb|DDash|, \verb|DDash2|, \verb|DDashC|;
\item Double dotted lines \verb|DDots|, \verb|DDots2|, \verb|DDotsC|;
\item Double zigzag lines \verb|DZigzag|, \verb|DZigzag2|, \verb|DZigzagC|;
\item Double photon lines \verb|DPhoton|, \verb|DPhoton2|, \verb|DPhotonC|;
\item Fermion-zigzag lines \verb|FZigzag|, \verb|FZigzag2|, \verb|FZigzagC|;
\item Fermion-photon lines \verb|FPhoton|, \verb|FPhoton2|, \verb|FPhotonC|;
\item Fermion-gluon lines \verb|FGluon|, \verb|FGluon2|, \verb|FGluonC|.
\end{itemize}

Other new features:
\begin{itemize}
\item Control variables \verb|DotsB|, \verb|DotsE| for dotted lines;
\item Polygonal vertices \verb|PVert| with control parameters \verb|VertA|, \verb|VertR|;
\item Dots on lines \verb|LDot|.
\end{itemize}

Each subroutine which draws something can be tuned by some parameters.
The full list of subroutines and parameters is given in the table on the last page;
it shows which parameters can influence the behavior of which subroutine.
It replaces a similar table in~\cite{Grozin:2022fde}.
there's one minor point not reflected there:
if a subroutine \verb|Foo2 x1 y1 x y| (where \verb|Foo| is some line kind)
is called with the end point $(x,y)$ coinciding with the current point,
it calls \verb|FooC|;
then the parameter \verb|CircleD| influences its behavior ---
it determines the rotation direction
(and hence the behavior of \verb|Arrow| and \verb|Mom| for this line).

The last line but one shows for which parameters \verb|Foo|
the variables \verb|DefFoo| are defined;
they contain the default values for \verb|Foo| which can be used
for restoring the default after some changes.
Variables \verb|Foo| whose default values are somehow trivial (e.g., 0 or 1)
don't have \verb|DefFoo|.
The last line shows variables which are automatically reset to defaults
after any call of the corresponding subroutines.
Those variables which define some ``style'' of some kind of lines are not reset automatically,
so, such a ``style'' can be re-used many times.
But some fine-tuning is most probably useful for the current line only,
it seems improbably that they will be useful for the next line.
Then the parameter is reset to default.

For using \verb|feyn.gle| a user should read~\cite{Grozin:2022fde}
and the current paper
(and, maybe, the \verb|GLE| manual).
The current version can be downloaded from \verb|GitHub|.
All \verb|.gle| scripts for the examples in the current paper
are attached to it.

\bibliographystyle{elsarticle-num}
\bibliography{feyn-1.2}

\newpage
\small
$\strut$\hspace{-1cm}
\setlength{\tabcolsep}{4pt}
\begin{tabular}{|l|ccc|c|cc|ccc|ccccc|cc|cc|cccc|ccc|c|}
\hline
&
\rotatebox{90}{\texttt{Gap}} &
\rotatebox{90}{\texttt{GapL}} \rotatebox{90}{\texttt{GapA}} &
\rotatebox{90}{\texttt{GapC\$}} &
\rotatebox{90}{\texttt{lwidth}} &
\rotatebox{90}{\texttt{DoubleA}} &
\rotatebox{90}{\texttt{DoubleB}} \rotatebox{90}{\texttt{DoubleE}} &
\rotatebox{90}{\texttt{DashL}} \rotatebox{90}{\texttt{DashF}} &
\rotatebox{90}{\texttt{DashB}} \rotatebox{90}{\texttt{DashE}} &
\rotatebox{90}{\texttt{DashN}} &
\rotatebox{90}{\texttt{DotR}} &
\rotatebox{90}{\texttt{DotsL}} &
\rotatebox{90}{\texttt{DotsB}} \rotatebox{90}{\texttt{DotsE}} &
\rotatebox{90}{\texttt{DotsN}} &
\rotatebox{90}{\texttt{FillC\$}} &
\rotatebox{90}{\texttt{PhotonL}} \rotatebox{90}{\texttt{PhotonA}} &
\rotatebox{90}{\texttt{PhotonN}} &
\rotatebox{90}{\texttt{GluonW}} &
\rotatebox{90}{\texttt{GluonS}} &
\rotatebox{90}{\texttt{DashPattern}} \rotatebox{90}{\texttt{DotsPattern}} &
\rotatebox{90}{\texttt{PatternB}} \rotatebox{90}{\texttt{PatternE}} &
\rotatebox{90}{\texttt{PatternN}} &
\rotatebox{90}{\texttt{DotsPatternB}} \rotatebox{90}{\texttt{DotsPatternE}} &
\rotatebox{90}{\texttt{VertA}} \rotatebox{90}{\texttt{VertR}} &
\rotatebox{90}{\texttt{ArrowL}} \rotatebox{90}{\texttt{ArrowA}} &
\rotatebox{90}{\texttt{MomL}} \rotatebox{90}{\texttt{MomD}} &
\rotatebox{90}{\texttt{CircleD}}\\
\hline
\verb|Feyn|  & \checkmark & \checkmark & \checkmark &&&&&&&&&&&&&&&&&&&&&&&\\
\verb|Feyn2| & \checkmark & \checkmark & \checkmark &&&&&&&&&&&&&&&&&&&&&&&\\
\verb|FeynC| & \checkmark & \checkmark & \checkmark &&&&&&&&&&&&&&&&&&&&&&& \checkmark \\
\hline
\verb|Fermion|  & \checkmark & \checkmark & \checkmark & \checkmark &&&&&&&&&&&&&&&&&&&&&&\\
\verb|Fermion2| & \checkmark & \checkmark & \checkmark & \checkmark &&&&&&&&&&&&&&&&&&&&&&\\
\verb|FermionC| & \checkmark & \checkmark & \checkmark & \checkmark &&&&&&&&&&&&&&&&&&&&&& \checkmark \\
\hline
\verb|Double|  & \checkmark & \checkmark & \checkmark & \checkmark & \checkmark & \checkmark &&&&&&&&&&&&&&&&&&&&\\
\verb|Double2| & \checkmark & \checkmark & \checkmark & \checkmark & \checkmark & \checkmark &&&&&&&&&&&&&&&&&&&&\\
\verb|DoubleC| & \checkmark & \checkmark & \checkmark & \checkmark & \checkmark & \checkmark &&&&&&&&&&&&&&&&&&&& \checkmark \\
\hline
\verb|Dash|  & \checkmark & \checkmark & \checkmark & \checkmark &&& \checkmark & \checkmark & \checkmark &&&&&&&&&&&&&&&&&\\
\verb|Dash2| & \checkmark & \checkmark & \checkmark & \checkmark &&& \checkmark & \checkmark & \checkmark &&&&&&&&&&&&&&&&&\\
\verb|DashC| & \checkmark & \checkmark & \checkmark & \checkmark &&& \checkmark & \checkmark & \checkmark &&&&&&&&&&&&&&&&& \checkmark \\
\hline
\verb|DDash|  & \checkmark & \checkmark & \checkmark & \checkmark & \checkmark & \checkmark & \checkmark & \checkmark & \checkmark &&&&&&&&&&&&&&&&&\\
\verb|DDash2| & \checkmark & \checkmark & \checkmark & \checkmark & \checkmark & \checkmark & \checkmark & \checkmark & \checkmark &&&&&&&&&&&&&&&&&\\
\verb|DDashC| & \checkmark & \checkmark & \checkmark & \checkmark & \checkmark & \checkmark & \checkmark & \checkmark & \checkmark &&&&&&&&&&&&&&&&& \checkmark \\
\hline
\verb|Dots|  & \checkmark & \checkmark & \checkmark & \checkmark &&&&&& \checkmark & \checkmark & \checkmark & \checkmark & \checkmark &&&&&&&&&&&&\\
\verb|Dots2| & \checkmark & \checkmark & \checkmark & \checkmark &&&&&& \checkmark & \checkmark & \checkmark & \checkmark & \checkmark &&&&&&&&&&&&\\
\verb|DotsC| & \checkmark & \checkmark & \checkmark & \checkmark &&&&&& \checkmark & \checkmark & \checkmark & \checkmark & \checkmark &&&&&&&&&&&& \checkmark \\
\hline
\verb|DDots|  & \checkmark & \checkmark & \checkmark & \checkmark & \checkmark & \checkmark &&&& \checkmark & \checkmark & \checkmark & \checkmark & \checkmark &&&&&&&&&&&&\\
\verb|DDots2| & \checkmark & \checkmark & \checkmark & \checkmark & \checkmark & \checkmark &&&& \checkmark & \checkmark & \checkmark & \checkmark & \checkmark &&&&&&&&&&&&\\
\verb|DDotsC| & \checkmark & \checkmark & \checkmark & \checkmark & \checkmark & \checkmark &&&& \checkmark & \checkmark & \checkmark & \checkmark & \checkmark &&&&&&&&&&&& \checkmark \\
\hline
\verb|Zigzag|  & \checkmark & \checkmark & \checkmark & \checkmark &&&&&&&&&&& \checkmark & \checkmark &&&&&&&&&&\\
\verb|Zigzag2| & \checkmark & \checkmark & \checkmark & \checkmark &&&&&&&&&&& \checkmark & \checkmark &&&&&&&&&&\\
\verb|ZigzagC| & \checkmark & \checkmark & \checkmark & \checkmark &&&&&&&&&&& \checkmark & \checkmark &&&&&&&&&& \checkmark \\
\hline
\verb|FZigzag|  & \checkmark & \checkmark & \checkmark & \checkmark &&&&&&&&&&& \checkmark & \checkmark &&&&&&&&&&\\
\verb|FZigzag2| & \checkmark & \checkmark & \checkmark & \checkmark &&&&&&&&&&& \checkmark & \checkmark &&&&&&&&&&\\
\verb|FZigzagC| & \checkmark & \checkmark & \checkmark & \checkmark &&&&&&&&&&& \checkmark & \checkmark &&&&&&&&&& \checkmark \\
\hline
\verb|DZigzag|  & \checkmark & \checkmark & \checkmark & \checkmark & \checkmark & \checkmark &&&&&&&&& \checkmark & \checkmark &&&&&&&&&&\\
\verb|DZigzag2| & \checkmark & \checkmark & \checkmark & \checkmark & \checkmark & \checkmark &&&&&&&&& \checkmark & \checkmark &&&&&&&&&&\\
\verb|DZigzagC| & \checkmark & \checkmark & \checkmark & \checkmark & \checkmark & \checkmark &&&&&&&&& \checkmark & \checkmark &&&&&&&&&& \checkmark \\
\hline
\verb|Photon|  & \checkmark & \checkmark & \checkmark & \checkmark &&&&&&&&&&& \checkmark & \checkmark &&&&&&&&&&\\
\verb|Photon2| & \checkmark & \checkmark & \checkmark & \checkmark &&&&&&&&&&& \checkmark & \checkmark &&&&&&&&&&\\
\verb|PhotonC| & \checkmark & \checkmark & \checkmark & \checkmark &&&&&&&&&&& \checkmark & \checkmark &&&&&&&&&& \checkmark \\
\hline
\verb|FPhoton|  & \checkmark & \checkmark & \checkmark & \checkmark &&&&&&&&&&& \checkmark & \checkmark &&&&&&&&&&\\
\verb|FPhoton2| & \checkmark & \checkmark & \checkmark & \checkmark &&&&&&&&&&& \checkmark & \checkmark &&&&&&&&&&\\
\verb|FPhotonC| & \checkmark & \checkmark & \checkmark & \checkmark &&&&&&&&&&& \checkmark & \checkmark &&&&&&&&&& \checkmark \\
\hline
\verb|DPhoton|  & \checkmark & \checkmark & \checkmark & \checkmark & \checkmark & \checkmark &&&&&&&&& \checkmark & \checkmark &&&&&&&&&&\\
\verb|DPhoton2| & \checkmark & \checkmark & \checkmark & \checkmark & \checkmark & \checkmark &&&&&&&&& \checkmark & \checkmark &&&&&&&&&&\\
\verb|DPhotonC| & \checkmark & \checkmark & \checkmark & \checkmark & \checkmark & \checkmark &&&&&&&&& \checkmark & \checkmark &&&&&&&&&& \checkmark \\
\hline
\verb|Gluon|  & \checkmark & \checkmark & \checkmark & \checkmark &&&&&&&&&&& \checkmark && \checkmark & \checkmark &&&&&&&&\\
\verb|Gluon2| & \checkmark & \checkmark & \checkmark & \checkmark &&&&&&&&&&& \checkmark && \checkmark & \checkmark &&&&&&&&\\
\verb|GluonC| & \checkmark & \checkmark & \checkmark & \checkmark &&&&&&&&&&& \checkmark && \checkmark & \checkmark &&&&&&&& \checkmark \\
\hline
\verb|FGluon|  & \checkmark & \checkmark & \checkmark & \checkmark &&&&&&&&&&& \checkmark && \checkmark & \checkmark &&&&&&&&\\
\verb|FGluon2| & \checkmark & \checkmark & \checkmark & \checkmark &&&&&&&&&&& \checkmark && \checkmark & \checkmark &&&&&&&&\\
\verb|FGluonC| & \checkmark & \checkmark & \checkmark & \checkmark &&&&&&&&&&& \checkmark && \checkmark & \checkmark &&&&&&&& \checkmark \\
\hline
\verb|Vert|    &&&& \checkmark &&&&&& \checkmark &&&& \checkmark &&&&&&&&&&&&\\
\verb|PVert|   &&&&&&&&&&&&&& \checkmark &&&&&&&&& \checkmark &&&\\
\verb|LDot|    &&&& \checkmark &&&&&& \checkmark &&&& \checkmark &&&&&&&&&&&&\\
\verb|Arrow|   &&&& \checkmark &&&&&&&&&&&&&&&&&&&& \checkmark &&\\
\verb|Mom|     &&&& \checkmark &&&&&&&&&&&&&&&&&&&& \checkmark & \checkmark &\\
\verb|Mom2|    &&&& \checkmark &&&&&&&&&&&&&&&&&&&& \checkmark & \checkmark &\\
\verb|DMom|    &&&& \checkmark & \checkmark &&&&&&&&&&&&&&&&&&& \checkmark & \checkmark &\\
\verb|DMom2|   &&&& \checkmark & \checkmark &&&&&&&&&&&&&&&&&&& \checkmark & \checkmark &\\
\hline
\textit{Def}       && \checkmark && \checkmark & \checkmark && \checkmark &&& \checkmark & \checkmark &&&& \checkmark && \checkmark &&&&&& \checkmark & \checkmark & \checkmark &\\
\textit{Autoreset} & \checkmark &&&&& \checkmark &&& \checkmark &&&& \checkmark &&& \checkmark &&&&& \checkmark &&&&&\\
\hline
\end{tabular}
\normalsize

\end{document}